# Infrared detection of aliphatic organics on a cometary nucleus

A. Raponi[1*], M. Ciarniello[1], F. Capaccioni[1], V. Mennella[2], G. Filacchione[1], V. Vinogradoff[3], O. Poch[4], P. Beck[4], E. Quirico[4], M. C. De Sanctis[1], L. Moroz[5,6], D. Kappel[7,5], S. Erard[8], D. Bockelée-Morvan[8], A. Longobardo[1,9], F. Tosi[1], E. Palomba[1], J.-P. Combe[10], B. Rousseau[1,4], G. Arnold[5], R. W. Carlson[11], A. Pommerol[12], C. Pilorget[13], S. Fornasier[8], G. Bellucci[1], A. Barucci[8], F. Mancarella[14], M. Formisano[1], G. Rinaldi[1], I. Istiqomah[4], C. Leyrat[8].

**Affiliations:**

[1]IAPS-INAF, Istituto di Astrofisica e Planetologia Spaziali, Area di Ricerca di Tor Vergata, Via del Fosso del Cavaliere, 100, 00133, Rome, Italy

[2]INAF – Osservatorio Astronomico di Capodimonte, Napoli, Italy

[3]Aix-Marseille Université, PIIM UMR-CNRS, 7345 Marseille, France

[4]Univ. Grenoble Alpes, CNRS, IPAG, 38000 Grenoble, France

[5]Institute for Planetary Research, German Aerospace Center (DLR), Rutherfordstraße 2, 12489 Berlin, Germany

[6]Institute of Earth and Environmental Science, University of Potsdam, Potsdam, Germany

[7]Institute of Physics and Astronomy, University of Potsdam, Potsdam, Germany

[8]LESIA, Observatoire de Paris, PSL Research University, CNRS, Sorbonne Universités, UPMC Univ. Paris 06, Univ. Paris Diderot, Sorbonne Paris Cité, 5 place Jules Janssen, 92195 Meudon, France

[9]DIST, Università Parthenope, Centro Direzionale ISola C4, 80143, Napoli, Italy

[10]Bear Fight Institute, Winthrop, WA, USA

[11]NASA JPL – Jet Propulsion Laboratory, California Institute of Technology, Pasadena, CA, USA

[12]Physikalisches Institut, Sidlerstr. 5, University of Bern, CH-3012 Bern, Switzerland

[13]Institut d'Astrophysique Spatial CNRS - Centre National de la Recherche Scientifique, Orsay, France

[14]Dipartimento di Matematica e Fisica "E. De Giorgi", Università del Salento, Via per Arnesano, Lecce, Italy

*Correspondence to: andrea.raponi@inaf.it

The ESA Rosetta mission[1] has acquired unprecedented measurements of comet 67P/Churyumov-Gerasimenko's (hereafter 67P) nucleus surface, whose composition, as determined by in situ and remote sensing instruments including VIRTIS (Visible, InfraRed and Thermal Imaging Spectrometer)[2] appears to be made by an assemblage of ices, minerals, and organic material[3]. We performed a refined analysis of infrared observations of the nucleus of comet 67P carried out by the VIRTIS-M hyperspectral imager. We found that the overall shape of the 67P infrared spectrum is similar to that of other carbon-rich outer solar system objects suggesting a possible genetic link with them. More importantly, we are also able to confirm the complex spectral structure of the wide 2.8-3.6 µm absorption feature populated by fainter bands. Among these, we unambiguously identified the presence of aliphatic organics by their ubiquitous 3.38, 3.42 and 3.47 µm bands. This novel infrared detection of aliphatic species on a cometary surface has strong implications for the evolutionary history of the primordial solar system and give evidence that comets provide an evolutionary link between interstellar material and solar system bodies[4].

The refractory dust component analyzed in situ by COSIMA aboard Rosetta revealed a mixture of carbonaceous matter and anhydrous mineral phases[5], consistent with the composition of Chondritic Porous - Interplanetary Dust Particles (CP-IDPs) and Anhydrous Antarctic MicroMeteorites (AMMs)[5]. The elemental composition is essentially chondritic, and shares similarities with the macromolecular Insoluble Organic Matter (IOM) extracted from primitive chondrites[6], with the exception of carbon that greatly exceeds the abundance measured in carbonaceous chondrites[5].

A complex and diverse nature of the organic matter was suggested by the mass spectrometers of ROSINA, COSAC and Ptolemy, which investigated the semi-volatile compounds. The aromatic compound toluene seems relatively abundant[7].

The organic-rich nature of 67P has also been highlighted by the reflectance spectra measured by VIRTIS in the 1-4 µm range[8,9], which showed a dark and red surface[10] with a broad absorption at 2.8-3.6 µm compatible with a complex mixture of various types of C-H and/or O-H chemical groups, with contribution of N-H groups such as ammonium ion ($NH_4^+$). However, a firm identification of the exact compounds has remained challenging because of the spectral congestion resulting from several molecular vibrations, and because earlier data analyses were affected by a significant number of systematic artifacts, masking weak spectral features.

Here, we have refined the VIRTIS dataset quality by solving two distinct issues with the calibration: (i) minimization of the instrumental artifacts, and (ii) improvement of the radiometric accuracy (see Supplementary Figs. 1, 2, 3 and Methods section for a full description of the inflight calibration refinements). From these improved data we have calculated an average spectrum from a few million surface spectra allowing minimization of the noise. The analysis has taken into account the datasets acquired during the early comet mapping phase (August-September 2014) to reduce the thermal emission contribution. We obtained a spectrum of unprecedented quality for the comet 67P nucleus surface, revealing a complex spectral structure, made up of weaker and ubiquitous spectral features inside the broad absorption at 2.8-3.6 µm. From this spectrum we have removed the contribution of thermal emission, which affects the spectral range longward of 3.0 µm (see Supplementary Fig. 4 and Methods for further details), finally yielding the radiance factor (*I/F)* spectrum shown in Fig. 1A.

Individual sub-features can be unambiguously identified within the broad absorption band between 2.8 and 3.6 µm (Fig. 1B). The strongest ones are centered at 3.1, 3.3, 3.38, 3.42, 3.47 µm. Another absorption band is detected at ~4.03 µm (see Supplementary Fig. 5 and discussion therein). Longward of 4.15 µm, the prominent thermal emission along with its large variability, due to complex illumination conditions across the surface, prevents a reliable removal of artifacts and the reduction of the noise.

The whole broad 2.8-3.6 µm absorption feature can be affected by the presence of O-H stretching in compounds such as carboxylic acids, alcohols, phenols[9], O-H bearing minerals, and water ice. Although water ice cannot explain the sub-features discussed here, it can contribute to the short-wavelength part of the absorption band (see Supplementary Fig. 6 and Methods), suggesting that the measured average spectrum is compatible with a ubiquitous presence of micron-sized water ice grains. This scenario is consistent with the previous findings about temporal variability of the broad absorption: micron-sized grains of water ice have been identified as a result of sublimation-recondensation processes occurring during the diurnal cycle of comet 67P[11] and along the seasonal cycle[12].

The absorption bands at 3.38 µm and 3.42 µm can be firmly assigned to asymmetric C-H stretching modes of the methyl ($CH_3$) and methylene ($CH_2$) aliphatic groups, respectively. A third aliphatic C-H stretching band is present at 3.47 µm and is assigned to the symmetric modes of $CH_3$, and $CH_2$ groups[13]. In pure aliphatic hydrocarbons, two features can be distinguished for the symmetric stretching vibration of the $CH_3$ and $CH_2$ groups, at 3.48 and 3.5 µm, respectively[13], while in the case of 67P spectrum these two modes are blended. The latter condition occurs when the aliphatic groups are bound to perturbing groups such as aromatic molecules[14,15]. This is commonly observed in both synthetic and natural carbonaceous material, and in meteoritic sample spectra[18,38].

The strength of the features detected on the spectrum of the surface of 67P is peculiar: the asymmetric stretching mode of $CH_3$ (3.38 µm) appears more intense than the asymmetric stretching of $CH_2$ (3.42 µm) (Fig. 2). Such configuration is unusual for most of the other investigated extraterrestrial materials. However, the relative intensities of the aliphatic sub-features in the 67P spectrum are affected by superposition with a variety of overlapping individual bands associated to other compounds (e.g. O-H or N-H) which can affect the apparent $CH_2/CH_3$ band ratio. Taking this into account, the aliphatic features are compatible with those reported for the interstellar medium (ISM)[16,17] as well as for IOM extracted from primitive Carbonaceous Chondrites (CCs)[18, 19, 20], as reported in Fig. 2 and Supplementary Fig. 7. These similarities in the spectra indicate a possible origin of the organic material of comet 67P in ISM, while additional spectral features reveal specific components, which could be in common with some other solar system bodies, as discussed below.

The 3.3 µm sub-feature in the 67P spectrum could be compatible with the aromatic C-H stretching mode. This is the case for the primitive CCs such as Murchison and Bells[21], but also in Wild 2 grains, where a 3.3 µm feature is present along with the aliphatic features. However, in those cases, the feature is not intense because aromatic carbons are present in the same carbon grain population as the aliphatic one, as cross-linked aromatic rings with aliphatic carbons, having almost no free aromatic C-H. Conversely, the stronger 3.3 µm sub-feature in the 67P spectrum can suggest the presence of an additional population of carbon grains, such as Polycyclic Aromatic Hydrocarbons (PAHs), with free aromatic C-H stretching giving a relatively intense strength of this feature. Emission of aromatic materials, such as PAHs, has been observed in ISM and proto-stellar disks[22]. The observation of the 3.3 µm feature in the 67P spectrum could be an evidence of direct condensation of PAHs in the comet.

On the other hand, the 3.3 µm as well as the 3.1 µm sub-feature can also be attributed to N-H vibrations. The 3.1 µm band is compatible with the N-H stretching mode[9], and many carriers could contribute to it, such as $NH_4^+$. The latter is a typical product formed in interstellar ice analogs[23] and has been proposed as a carrier of the 3.1 µm band in the interstellar ices in dense regions or in the outer solar system[24,25]. The sub-features at 3.1, 3.3 and 3.4 µm, although previously never detected on a cometary surface, have been observed on other bodies in the outer solar system. The average spectrum of six Trojan asteroids shows the presence of the 3.1 µm absorption (Fig. 3) attributed to the N-H bond, and the tentative presence of the 3.3 µm and 3.4 μm bands[26]. Other bodies show the simultaneous presence of two of these three features: distinct 3.3 and 3.4 μm features, attributed to PAHs and aliphatics[26], are present on Saturn's moon Phoebe and in the organic material deposited on Iapetus from the Phoebe ring itself[27]. Asteroids of the outer main belt (24) Themis[28], (65) Cybele[29], (52) Europa[30], (361) Bononia[30], and the irregular Jupiter satellite Himalia[28] show the absorption at 3.1 μm and another shallower absorption in the 3.3 – 3.4 μm range (Fig. 3), even if no distinct aliphatic signatures can be recognized. We also note that, in addition to the presence of some of the 3.1, 3.3 and 3.4 μm sub-features, a few of these bodies also show a positive spectral slope[32] of the infrared reflectance continuum, which is similar to the one shown by 67P (Fig. 3). In this context, it is worth mentioning the case of Ceres, which shows a local presence of aliphatic carbons, together with the absorption at 3.1 μm attributed to $NH_4^+$[33]. However, other absorptions populate the infrared spectrum of this dwarf planet indicating a pervasive aqueous alteration of the surface, thus weakening the comparison with the pristine material of the comet[34].

Most of the bodies mentioned above have had a complex dynamical history, which brought them from the Kuiper belt to their present position after gravitational scattering or capture by giant planets[35]. A similar dynamical evolution is expected for Jupiter family comets, such as 67P[4]. For 67P, several evidences obtained by Rosetta, e.g. the D/H ratio reported by Altwegg et al.[36] or the low temperature (<30 K) of the nucleus formation derived by Rubin et al.[37], indicate that 67P must have had a common origin with other Kuiper belt objects[3]. Nonetheless, in view of its small size and mass, 67P has probably not experienced any aqueous alteration unlike the larger bodies mentioned in the previous paragraph, which could have modified their pristine composition.

Comets are believed to be chemically unaltered bodies[37] and as such they could reveal properties of the pristine organic material. In particular, given the continuous erosion due to comet activity, the surface of the 67P nucleus is continuously resupplied by unprocessed material[9]. The ubiquitous presence of the aliphatic matter now unambiguously identified in the 67P spectrum, and the relative intensities of the $CH_3$ and $CH_2$ asymmetric bands, which point to a $CH_3$-rich aliphatic linkage, raise the question about the origins of organic material in comets: either protosolar and/or interstellar. COSIMA results shed new light on the refractory matter of the 67P surface. The large amount of carbon[6] suggests its earlier presence in the solar nebula, probably inherited from the ISM and then preserved in the cold environment where the comet formed.

An interstellar origin for the aliphatic carbons observed in 67P is supported by laboratory experiments[38]: they have indicated that a single carbon grain population can be responsible for the aliphatic bands observed in diffuse interstellar clouds and in the solar system materials (Stardust particles, and CP-IDPs), as well as for the broad absorption detected in dense clouds of the ISM. This suggests an evolutionary connection of the organic material in these environments[38,39].

Similarly, the extent of interstellar inheritance in IOM has been largely debated. For instance, the abundance ratio of chondritic IOM to presolar nanodiamonds in chondrites is roughly constant, which is consistent with an interstellar origin for the IOM[21]. On the other hand, the aliphatic feature in the ISM is due to hydrogenated amorphous carbons that are devoid of oxygen[40] unlike the IOM. This can imply either a protosolar origin, or that some process (such as the asteroid hydrothermal alteration) have inserted oxygen as a heteroatom in the macromolecular structure of the IOM[21].

Analyses performed on grains from comet Wild 2 collected by Stardust[41] and CP-IDP could not provide an unequivocal interpretation: deuterium and nitrogen-15 excesses suggest that organics are compatible with an interstellar heritage and processing of ices[42]. Nonetheless, their spectral characteristics in the near IR are different with respect to the ISM spectra, in particular a higher $(CH_2/CH_3)_{asym}$ intensity ratio, suggesting also a protosolar origin[43]. However, intensities of these features could have been perturbed by the interaction of the particles with terrestrial environment, and with aerogel in the case of the Wild 2 grains[42].

In this context, the aliphatic features observed on the 67P spectrum are comparable to typical aliphatic ISM features suggesting a possible evolutionary link between hydrocarbons in diffuse ISM and comets. On the other hand, the aliphatic signatures on 67P are also compatible with those of the primitive IOM chondrites, suggesting a link in carbon composition among chondrites and comets. The spectrum of 67P also blurs the distinction between comets and other primitive objects of the outer solar system (Fig. 3), reinforcing the idea of a continuum between comets and asteroids.

The organic material inherited by comets and other bodies from the ISM was delivered to the planets of the inner solar system, including the Earth[44], possibly favoring prebiotic environments. Distinguishing between an interstellar or protosolar origin of the building blocks of life would have implications on their availability for our and other planetary systems in the universe.

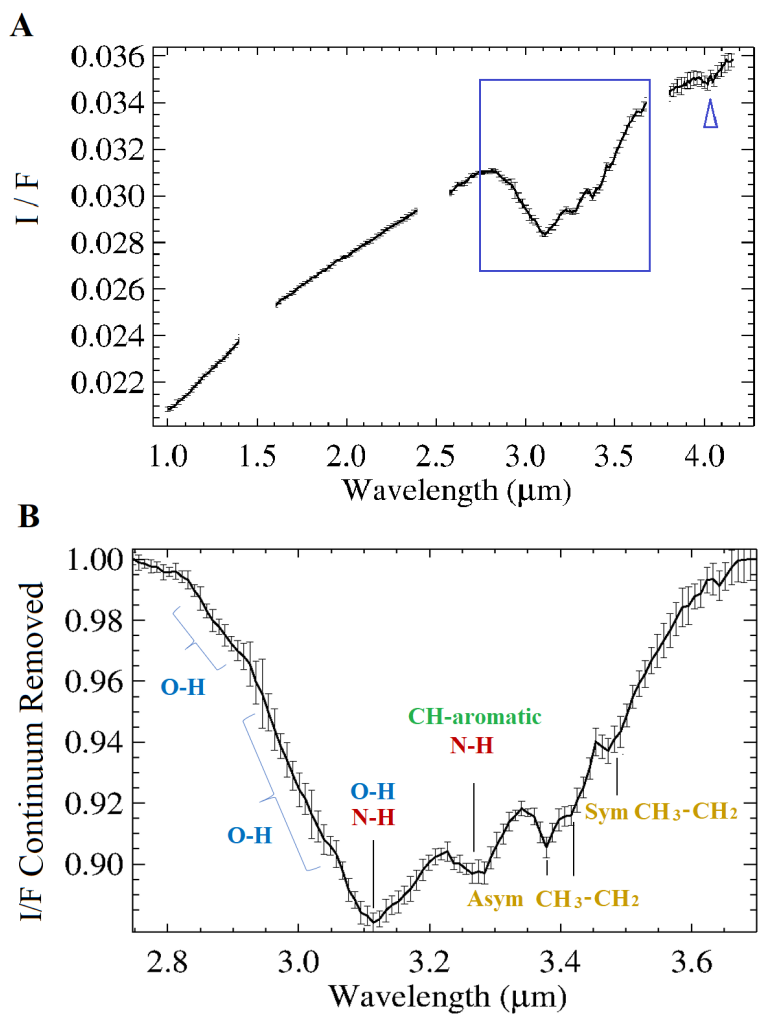

**Figure 1**. Panel A: calibrated average *I/F* spectrum of 67P after removal of thermal emission (see Methods). Blue triangle indicates a tentative absorption (see Fig. S5). Missing parts of the spectrum correspond to the junctions of the instrument's order sorting filters.

Panel B: continuum-removed average spectrum of 67P across the broad absorption at 2.8-3.6 µm, corresponding to the blue box of panel A. Attributions of the main spectral features are indicated. Estimation of the error bars is described in the Methods section.

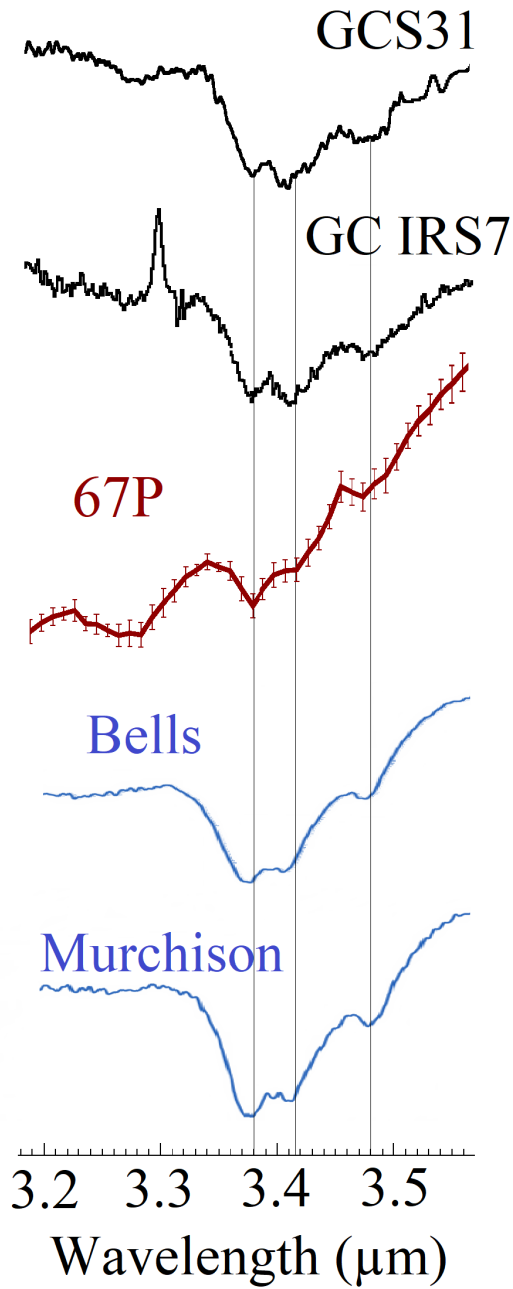

**Figure 2**. Comparison of the aliphatic features on 67P, ISM, and IOM, shown in arbitrary units, with offsets. Average *I/F* spectrum of 67P normalized to the continuum (red line with error bars). Two spectra of interstellar material acquired in the line of sight of the galactic center[17] (black curves), and two spectra of chondritic IOM (blue curves)[18]. Vertical lines indicate absorption features of aliphatics.

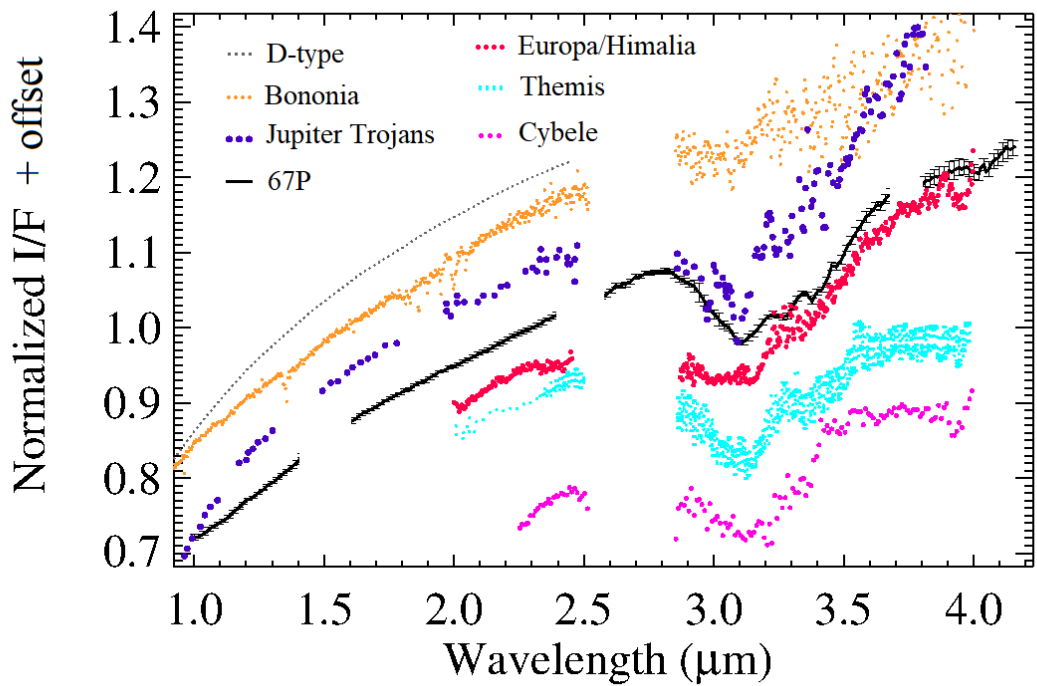

**Figure 3**: Comparison between 67P spectrum and other bodies of the solar system normalized at 2.3 μm. Arbitrary offsets were applied to the spectra, except for 67P. The absorption features centered at 3.1 μm and 3.3 - 3.4 μm are not only observed on comet 67P but also on several asteroids: (361) Bononia[30], (52) Europa[30], (24) Themis[28], (65) Cybele[29], and Jupiter Trojans (average spectrum of six from the "less-red" group)[26]. Himalia (not shown here) is an irregular satellite of Jupiter having a strikingly similar spectrum as (52) Europa, as shown in Brown & Rhoden[31]. Trojans and average D-type asteroids[32] (such as Bononia) show a similar red spectral slope with respect to that of comet 67P.

## Methods

**In-flight calibration of VIRTIS-M-IR**

The VIRTIS-M (Visual Infrared and Thermal Imaging Spectrometer – Mapper)[2] IR channel acquired hyperspectral images in the range 1 - 5 µm, across 432 spectral bands, and through 256 spatial samples of 250 µrad/px x 250 µrad/px.

During the cruise of the mission VIRTIS-M onboard Rosetta acquired spectra of the Earth, the Moon, Mars, asteroids Steins and Lutetia, and some point targets, prior to the rendezvous with comet 67P/Churyumov-Gerasimenko. The spectra of all these targets show a consistent pattern independently of the object observed. This is an indication that the measured spectrum is affected by systematic artifacts superimposed on the real spectral features, preventing their clear detection. In particular, two types of artifacts can be distinguished:
1) artifacts linearly dependent on the signal;
2) indentation between odd and even spectral bands, not correlated with the signal.
We developed an in-flight calibration refinement in order to minimize their effects, as follows:

Artifact removal:

We consider an average spectrum of the comet nucleus signal obtained from ~2.7 million spectra acquired during the first mapping phase of the Rosetta mission in August-September 2014, from the northern hemisphere and equatorial regions of the comet, and an average spectrum of the Lutetia asteroid, for each spatial sample (s) of the detector. We excluded from the average the spatial pixels corresponding to shadowed surface parts. Spectra are processed sample by sample to trace the variability of the artifacts across the focal plane. The ratio between 67P and Lutetia spectra allows removal of spectral artifacts while keeping information of the real features.

Assuming the spectrum of Lutetia is devoid of small real features (namely involving few spectral bands), we model it with a polynomial interpolation (*interp*) representing the absolute reference, which is then used to produce an artifact-removed (*AR*) spectrum of 67P (see Eq. 1):

(1) $$\frac{\frac{I}{F}(\lambda,s)_{67P}}{\frac{I}{F}(\lambda,s)_{Lutetia}} \cdot \frac{I}{F}(\lambda,s)_{\substack{interp \\ Lutetia}} = \frac{I}{F}(\lambda,s)_{\substack{AR \\ 67P}}$$

Here *I/F* is the radiance factor (π · bidirectional reflectance)

Then we calculated the average *I/F* and the standard deviation of the normalized *I/F* along the samples (*s*) to obtain respectively the artifact-removed (*AR*) spectrum and the uncertainties for each band (*λ*).

The resulting spectral features on the average calibrated spectrum of 67P can be also obtained using data from Mars observations in the range 3.1 – 3.5 μm, confirming the reliability of the result obtained.

As a further check we performed the ratio between two spectra of the internal calibration lamp: one acquired during the 67P mapping phase and the other during the Lutetia flyby. Since no significant variations were found between the two acquisitions, we conclude that the detector has not experienced any aging or contamination effects between Lutetia observations (used as a reference for the calibration) and the 67P encounter.

Odd-even bands indentation removal

The average spectrum of the comet, cleaned from artifacts, still presents a source of non-Poissonian noise introduced by the electronics of the instrument. Due to the detector's architecture, the even spectral channels' response is affected by the temperature of the instrument, which introduces spurious offsets along the wavelengths especially at low fluxes (see Fig. S1). Thus, we replaced the signals of the even channels by an average of the contiguous odd spectral channels. Although this implies a halving of the spectral sampling, it only produces a marginal reduction of the spectral resolution, given that the nominal spectral resolution is ~15 nm FWHM and the final spectral sampling is 19.4 nm[45], while it improves the noise level in a way to clearly detect the small features which are the subject of this work. A possible approach to remove these odd-even artifacts in a situation with more favorable data coverage was detailed by Kappel et al.[46] for the sister instrument VIRTIS-M-IR on Venus Express.

Absolute calibration with star observations

A new version of the radiometric calibration is performed using stellar sources. Both VIRTIS-Rosetta and VIMS-Cassini observed stars during the cruise phase of the respective missions. This allows the comparison of the fluxes observed by both instruments to perform an inter-calibration (see Fig. S2). In particular, we compared two acquisitions of Arcturus performed by VIRTIS-M-IR with six observations performed by VIMS onboard Cassini (from Cassini Atlas of Stellar Spectra[47]). The ratio of the average fluxes observed by the two instruments provides a correction factor as a function of wavelength, which can be applied to the whole VIRTIS-M-IR dataset (see Fig. S2). We rely on VIMS spectra because of their consolidated calibration[48], which has been thoroughly tested and improved over the duration of the Cassini mission.

**Thermal emission removal**

The average calibrated spectrum as derived according to the previous section is affected by thermal emission of the surface of the comet. In particular, it affects the shape of the spectrum longward of 3 µm, with increasing effect for longer wavelength. We thus removed the thermal emission similarly to Raponi et al.[49]. The total signal is modeled as the sum of the thermal emission (a gray-body Planck emission, where temperature and emissivity are free parameters, with emissivity constant with wavelength) and the solar light reflected from the surface. The latter is modeled as a linear function with slope and intercept as free parameters. The linear function is used as a reference in order to remove the continuum of the absorption as shown in Fig. 1B. The spectral fit is performed from 2.2 µm longward, excluding the spectral range of the absorption at 2.8 – 3.6 µm, to retrieve free parameters. Once the total signal is modeled, the thermal emission is subtracted from the measured calibrated signal (Fig. S4).

**Modeling of possible contribution of water ice to the average spectrum of the 67P nucleus**

Water ice has been observed in localized regions of comet 67P in the form of both small grains and patches with larger grains[11,49]. The processes involving water ice are outside the interest of the present work, since here we are focused on the average spectrum of the comet nucleus during the first phase of the mission when such processes were weakly active or very localized. However, it is possible that even the most dehydrated regions of the comet contained a small fraction of icy grains.

We performed modeling of the *I/F* average spectrum of comet 67P to evaluate the possible contribution of water ice to the broad absorption in the range from 2.8 to 3.6 µm. We used the Hapke IMSA radiative transfer model[50] to simulate the resulting reflectance of an intimate mixture of a dark-red terrain and water ice. The bidirectional reflectance (*r*) is given by Eq. 2,

(2) $$r = \frac{SSA}{4\pi} \frac{\mu_0}{\mu+\mu_0} [p(g) + H(SSA, \mu)H(SSA, \mu_0) - 1] \times S(i, e, g, \theta),$$

where *i, e, g* are the effective incidence, emission, and phase angles, respectively, and $\mu_0$, $\mu$ are the cosines of the effective incidence and emission angles. The parameters that contain most of the spectral information are the single scattering albedo (*SSA*), and the related Ambartsumian–Chandrasekhar functions *H(SSA, µ)* describing the multiple scattering components. Other terms that describe the photometric behavior as a function of viewing geometry are: the single particle phase function *p(g)*; the shadowing function modeling large-scale roughness *S(i, e, g, θ)*, with *θ* being the average surface slope.

These photometric parameters are defined after Ciarniello et al.[10]. The SSA has been modeled as an intimate mixture of water ice (*W*) and a simulated dark-red endmember (*D*), which implies that the particles of the endmember materials are in contact with each other and are both involved in the scattering of a single photon. The $SSA_W$ of water ice is defined starting from the grain size and the optical constants, as given by the equivalent-slab model[47]. We used the optical constants from Warren and Brandt[48]. The average $SSA_D$ of the dark-red endmember is modeled with a second-degree polynomial function. The weight $p_D$ and $p_W$ of the two endmembers in the mixture are defined as the corresponding cross section fraction with respect to the total cross section of the grains as seen by the photon propagating in the medium. Hereafter we refer to the weights $p_D$ and $p_W$ as "abundances".

(3) $\quad SSA = SSA_D\, p_D + SSA_W\, p_W \quad$ with: $p_D + p_W = 1$

Free model parameters to be retrieved by the fitting routine are:
- Coefficients of the second-degree polynomial function which define the dark-red endmember;

- Temperature and effective emissivity which model the thermal emission (as discussed in the previous section).

We produced simulated spectra by varying (i) water ice abundance (see Fig. S6, panel A); (ii) grain size of the water ice (see Fig. S6, panel B).

Resulting simulated spectra show that water ice grains larger than ~5 µm are incompatible with the measured 67P average spectrum, both because of the mismatch of the absorption band at ~3 µm and the emergence of the absorption bands at 1.5 and 2.0 µm which are not observed.

Simulated spectra obtained by fixing the water ice grain size to 0.5 µm, and by varying the abundance, show that a small amount of water ice in the order of 2% can contribute to the measured absorption. However, modeling of reflectance spectra of intimate mixtures involving sub-micrometer grains violates the applicability of Hapke theory and should be regarded as a qualitative result.

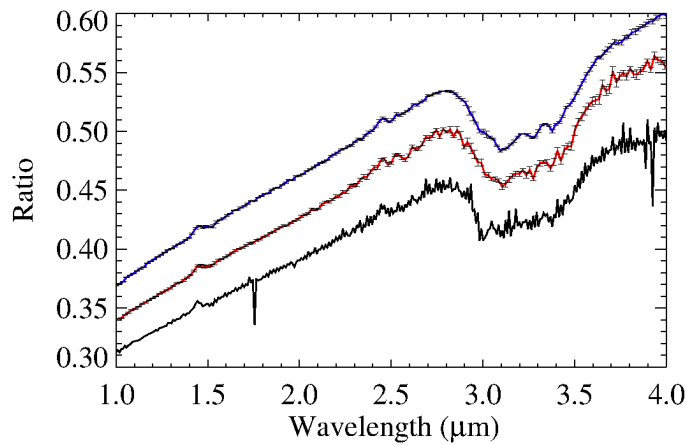

**Supplementary Figure 1**. Black line: ratio of the average *I/F* spectrum of 67P over the average *I/F* spectrum of (21) Lutetia calculated for one single sample of the spectrometer slit. Red and blue lines are the average ratio along the slit considering only the odd (+0.03 offset) and even (+0.06 offset) spectral channels, respectively. The error bars are the standard deviations calculated over the whole spectrometer slit. At 1.5 and 2.5 µm the junctions of the instrument's order sorting filters produce unreliable features.

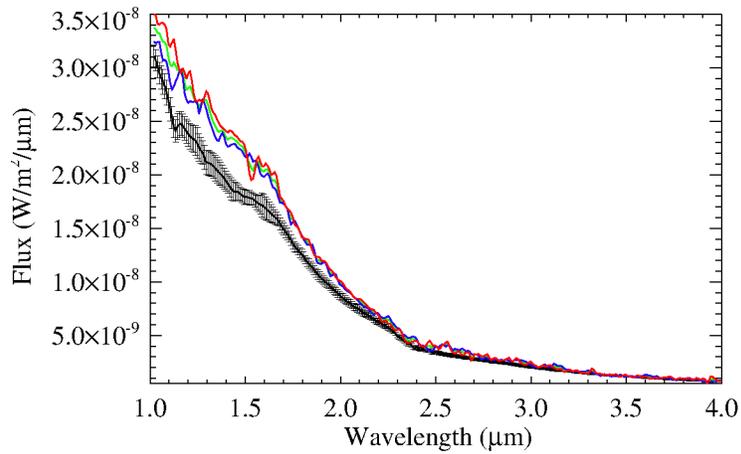

**Supplementary Figure 2**. Average flux of the star Arcturus acquired by VIMS-Cassini (black line). Error bars are the standard deviations over six observations. Flux of Arcturus acquired by VIRTIS-M-IR (red and blue line) and their average (green line) before absolute calibration with star observations.

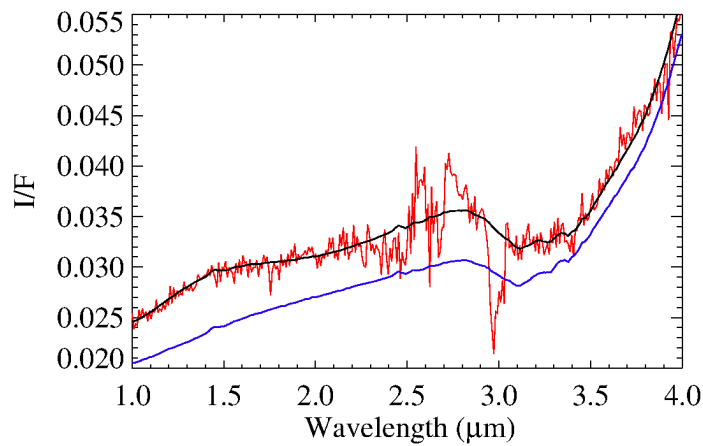

**Supplementary Figure 3**. Average spectrum of comet 67P from a single sample of the spectrometer slit (red). The calibrated spectrum after artifact and odd-even effect removal (black). Final shape of the spectrum after application of the stellar calibration factor (blue). No offset is added to the spectra. At 1.5 and 2.5 μm the junctions of the instrument's order sorting filters produce unreliable features.

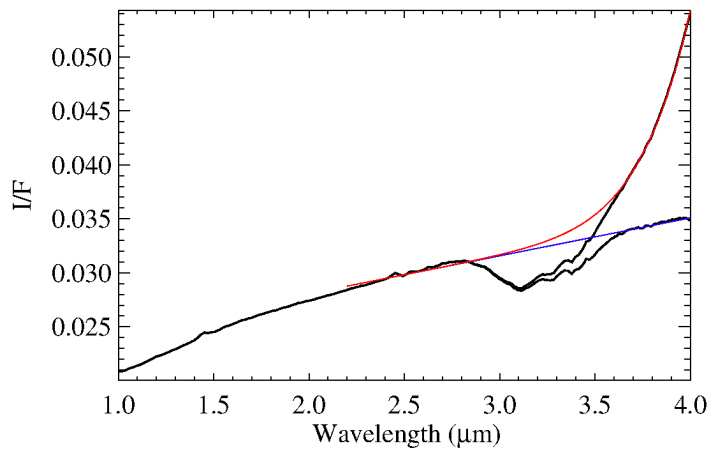

**Supplementary Figure 4**. The total modeled signal (red line) is the sum of the modeled reflectance continuum (blue line) and the thermal emission. The latter is subtracted from the measured calibrated signal. At 1.5 and 2.5 µm the junctions of the instrument's order sorting filters produce unreliable features.

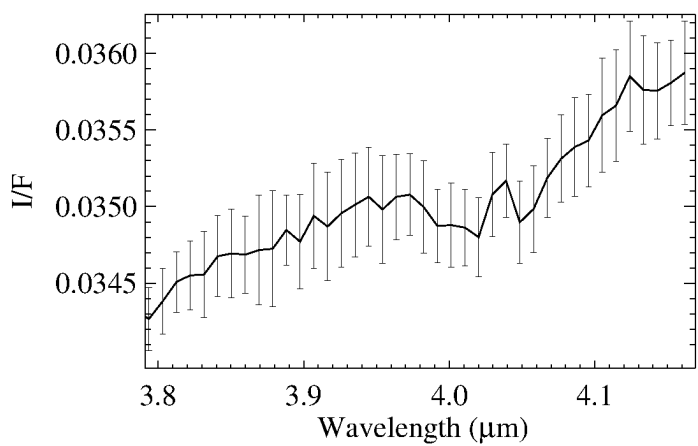

**Supplementary Figure 5**. The calibrated 67P spectrum after thermal emission removal shows a tentative absorption centered at 4.03 µm (blue triangle in Fig. 1A), or a tentative doublet at 4.01

and 4.05 µm, which could be attributed to carbonates. The formation of carbonates following aqueous alteration is ruled out for 67P. Toppani et al.[52] suggested that carbonates can be grown from the non-equilibrium condensation of a silicate gas in a $H_2O$-$CO_2$-rich vapor. However, a clear attribution to the 4.03 µm feature remains to be established.

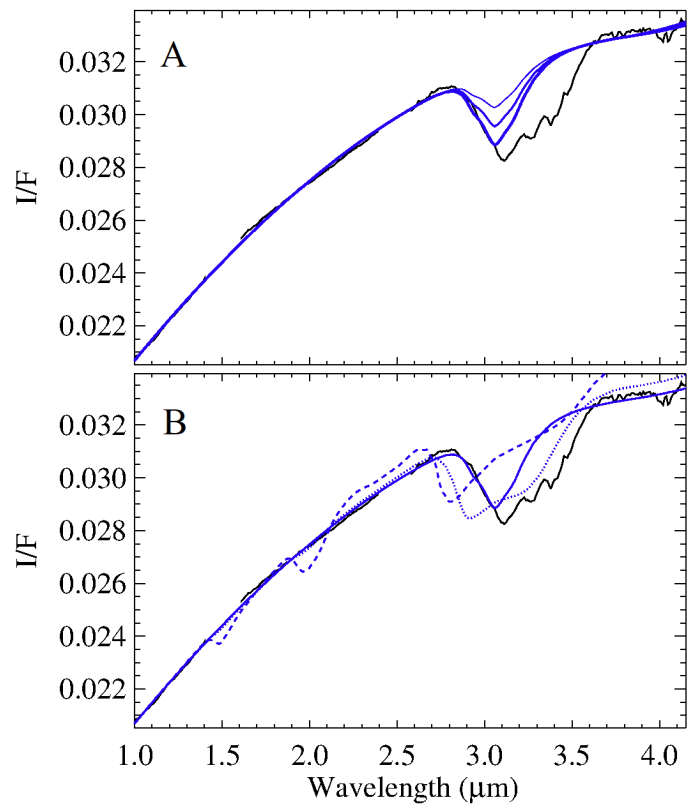

**Supplementary Figure 6**. Black line: average 67P spectrum; blue lines: modeled water ice mixed with the dark-red endmember. Panel A: water ice grain size fixed at 0.5 µm; abundance fixed at 1%, 1.5%, and 2% (respectively solid line with increasing thickness). Panel B: water ice abundance fixed at 2%; grain sizes fixed at 0.5 µm (solid line), 5 µm (dotted line), 50 µm (dashed line). We remark that the simulation with particles 0.5 µm-sized is not strictly compatible with a rigorous

application of the Hapke model which is based on the geometric optic assumption. Nevertheless, the trend with grain size still points to the possible presence of submicron grains.

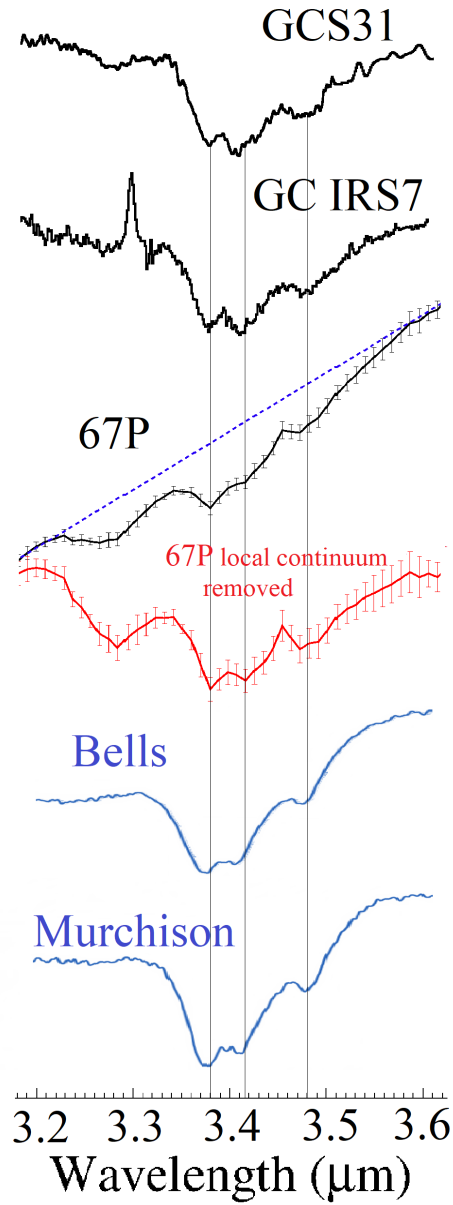

**Supplementary Figure 7**. Aliphatic features in 67P spectrum are affected by superposition with overlapping bands associated to other compounds (e.g. O-H or N-H). Assuming the contribution of the latter in correspondence to the range 3.2 – 3.6 μm is approximated with the blue dashed line,

the effective shape of the aliphatic features can be estimated by removing this "local" continuum. This plot, similar to Fig. 2 in the main text, shows a comparison of the aliphatic features on 67P, ISM, and IOM, in arbitrary units, with offsets: average I/F spectrum of 67P (black line with error bars) and local continuum (blue dashed line); spectrum of 67P normalized to the local continuum (red line); two spectra of interstellar material acquired in the line of sight of the galactic center[17] (black curves), and two spectra of chondritic IOM (blue curves)[18]. Vertical lines indicate absorption features of aliphatics.

**Acknowledgments:** We thank the following institutions and agencies for supporting this work: Italian Space Agency (ASI, Italy) contract number I/024/12/2, Centre National d'Etudes Spatiales (CNES, France), DLR (Germany), and NASA (USA) Rosetta Program. VIRTIS was built by a consortium, which includes Italy, France, and Germany, under the scientific responsibility of the Istituto di Astrofisica e Planetologia Spaziali of INAF, Italy, which also guides the scientific operations. The VIRTIS instrument development, led by the prime contractor Leonardo Company (Florence, Italy), has been funded and managed by ASI, with contributions from Observatoire de Meudon financed by CNES, and from DLR. We thank the Rosetta Science Ground Segment and the Rosetta Mission Operations Centre for their support throughout all the phases of the mission. The VIRTIS-calibrated data are available through the ESA's Planetary Science Archive web site http://www.rssd.esa.int/index.php?project=PSA&page=index. This work takes advantage of the collaboration of the ISSI international team "Comet 67P/Churyumov-Gerasimenko Surface Composition as a Playground for Radiative Transfer Modeling and Laboratory Measurements", number 397. Dr. Driss Takir is acknowledged for providing the reflectance spectra of (52) Europa and (361) Bononia. L. M. acknowledges the DFG (Deutsche Forschungsgemeinschaft) grant MO 3007/1-1. D. K. acknowledges DFG-grant KA 3757/2-1. P.B. acknowledges funding from the H2020 European Research Council (ERC) (SOLARYS ERC-CoG2017_771691). This research has made use of NASA's Astrophysics Data System Service.

**Author contributions:** A.R. wrote the manuscript, calibrated the data, and performed data analysis and interpretation; M. C., F. C., V. M, G. F., V. V., P. B., E. Q., M. C. D., and L. M. contributed to interpretation; G. F. contributed to calibration of the data; all co-authors helped with manuscript preparation.

**Competing interests:** The authors declare that they have no competing interests.

**Data and materials availability:** All data needed to evaluate the conclusions in the paper are present in the paper and/or the Supplemental Information.

The VIRTIS calibrated data are available through ESA's Planetary Science Archive (PSA) (http://www.cosmos.esa.int/web/psa/rosetta).